\documentclass[prd,aps,twocolumn,superscriptaddress,amsmath,floatfix,amssymb]{revtex4}
\usepackage{graphicx}

\newcommand{\beqn}{\begin{eqnarray}}
\newcommand{\eeqn}{\end{eqnarray}}
\newcommand{\eq}[1]{(\ref{#1})}

\newcommand{\Tr}{ {\rm Tr} \, }

\newcommand{\sign}{ {\rm sign} \,  }
\newcommand{\logo}{\\ \vskip -15mm \leftline{\includegraphics[scale=0.3,clip=false]{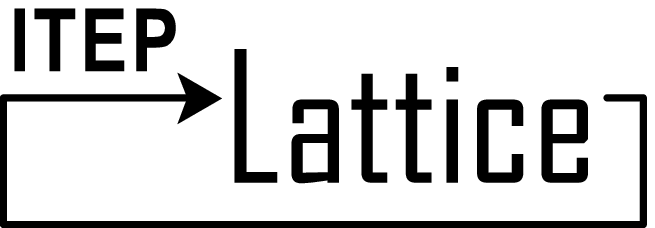}} \vskip 7mm}

\newcommand{\fm}{\text{\fontsize{10pt}{14pt}fm}}
\newcommand{\Mev}{\text{\fontsize{10pt}{14pt}MeV}}
\newcommand{\Gev}{\text{\fontsize{10pt}{14pt}GeV}}

\begin{document}
\sloppy

\title{The $\rho$ and $A$ mesons in strong abelian magnetic field in SU(2) lattice gauge theory.\logo}

\author{E.V. Luschevskaya}

\author{O.V. Larina}
\affiliation{ITEP, B. Cheremushkinskaya 25, Moscow, 117218 Russia}

\begin{abstract}
We calculated correlators of vector, axial and pseudoscalar currents  in external strong abelian magnetic field according to  $SU(2)$ gluodynamics.
The masses of neutral $\rho$ and $A$ mesons with various spin projections to the axis parallel to the external magnetic field $B$ have been calculated. We found that the masses of  neutral mesons with zero spin $s=0$ decrease in increasing
 magnetic field, while the masses of the $\rho$ and $A$ mesons with  spin $s=\pm 1$ increase
 in the mentioned field. Also we performed extrapolation and renormalization of masses on the lattice.

\end{abstract}

\maketitle

\section{Introduction}

 Quantum Chromodymanics in  external magnetic field is an area that presents enormous interest for physicists.
Strong magnetic fields could be associated with  formation of the early Universe.
It is assumed that magnetic fields  $\sim 2$ GeV existed in the  Universe during the electroweak phase transition~\cite{Hector:00}.
It is known that cosmic space objects called magnetars or neutron stars possess magnetic field in their cores  equal to  $\sim 1\ \Mev$.

Recently there was a number of amazing effects experimentally discovered and theoretically approved.
For instance, STAR collaboration detected chiral magnetic effect during non-central collisions of gold ions based on data provided by the RHIC  \cite{Voloshin:04:1,Voloshin:08:1,Kharzeev:08:1}.
 Later this effect was also observed in ALICE experiment  at the LHC.

 The values of  magnetic fields in  non-central heavy-ion collisions can reach up to  $15 m^2_{\pi} \sim 290\, \Mev^2$ ~\cite{Skokov:2009}, energy  of  hadronic scale.
This magnetic field is caused by motion of  ions,  and it creates charge asymmetry of particles emitted from  different sides of the reaction plane.
 Such  strong magnetic field can be created in terrestrial laboratories, which  makes it possible to explore quark - hadronic matter under  extreme conditions.

Strong enough magnetic fields  lead also to  modifications of the QCD phase diagram.
Phenomenological models show that the critical temperature of  transition between  phases of confinement and deconfinement varies with the increase of  external magnetic field, and the phase transition becomes the one of the first order~\cite{Mizher:2010}.
The increase of $T_c$ is also predicted in Nambu-Jona-Lasinio models: NJL, EPNJL, PNJL \cite{Gatto:2010} and PNJL$_8$ \cite{Kashiva:2011}, Gross-Neveu model \cite{Kanemura:1997, Klimenko:1992}.

The first prediction of  lattice simulations with two flavours in QCD regarding  behavior of the deconfinement temperature and the chiral symmetry restoration in increasing  magnetic field was made in \cite{Massimo.DElia:2010}.
However lattice simulations  with $N_f=2+1$  \cite{Bali:2011} revealed that $T_c$ decreases under  increasing
field value.
As a matter of fact, the chiral perturbation theory  predicts the decrease of  the transition temperature
under increasing  magnetic field value as well ~\cite{Agasian:2008}.

These effects were studied in the past but only  for
the case of a chromomagnetic (not usual abelian magnetic) field  \cite{Paolo.Cea:2002,Paolo.Cea:2005,Paolo.Cea:2007}.
  External magnetic fields lead to the enhancement of the chiral symmetry breaking ~\cite{Gusynin:1996,Klevansky:1989,
Ebert:1999,  Fraga:2008,Goyal:1999}. The analytical calculations ~\cite{Agasian:1999,Cohen:2007,Smilga:1997} predict a linear increase of the chiral condensate under increasing magnetic field in the leading order. Lattice simulations show that  chiral condensate
depends on the strength of applied field as exponent function with  $n = 1.6 \pm 0.2$
~\cite{Braguta:2010}. AdS/CFT approach shows the quadratic behaviour~\cite{Zayakin:2008}.

In the framework of the Nambu-Jona-Lasinio model it was shown that QCD vacuum becomes a superconductor \cite{Chernodub:2010,superconductivity} along the direction of the magnetic field  in the presence of  sufficiently strong magnetic fields
($B_c=m^2_{\rho}/e \simeq 10^{16}$ Tl). This transition to  superconducting phase is accompanied by  condensation of  charged $\rho$ mesons.  Calculations on the lattice ~\cite{Braguta:2011}  indicate  existence of the superconducting phase as well.
We have investigated the behavior of masses of neutral $\rho$ and $A$ mesons with  spin projection $s = 0,\, \pm 1$ to the axis of the magnetic field. In QCD condensation of neutral $\rho$ and $A$ mesons in external magnetic field would become an evidence of the superfluid phase existence. In \cite{Simonov:2013} they calculated the mass of neutral vector $\rho$ meson
according to the relativistic quark-antiquark model and found out that in the phase of confinement the mass of this neutral $\rho$ meson with zero spin increases under raising magnetic field which contradicts the results presented
in \cite{Chernodub:2010}.
This paper is organized in the following way.
In Section~\ref{Setup} we describe  technical and numerical specifications of our simulations.
In Section~\ref{Observables} we discuss  measured observables. Section~\ref{Methods} is devoted to the methods of calculations.
The results of our calculations are presented in Sections~\ref{Results1} and ~\ref{Results2}.

\section{Details of the calculations}
\label{Setup}
To generate $SU(2)$ gauge field configurations we use the tadpole improved Symanzik action
\begin{equation}
S=\beta_{imp} \sum_{pl} S_{pl}-\frac{\beta_{imp}}{20 u^2_0}\sum_{rt}S_{rt},
\label{action}
\end{equation}
where $S_{pl,rt}=(1/2)\Tr(1-U_{pl,rt})$ is the plaquette (denoted by $pl$) or 1$\times$2 rectangular loop term ($rt$),
$u_0=(W_{1\times1})^{1/4}=\langle(1/2)\Tr U_{pl}\rangle^{1/4}$ is the input tadpole factor computed at zero temperature \cite{Bornyakov:2005}.
This action  suppresses  ultraviolet dislocations, leading to  non-physical near-zero modes of the Wilson-Dirac operator.

We calculated fermionic spectrum in the presence of $SU(2)$ gauge fields using the chiral-invariant overlap operator proposed by Neuberger ~\cite{Neuberger:1997}. This operator allows to explore the theory without chiral symmetry breaking and can be written as follows:
\begin{equation}
D_{ov}=\frac{\rho}{a}\left( 1+D_W/\sqrt{D^{\dagger}_W D_W}  \right),
\label{overlap}
\end{equation}
where $D_W=M-\rho/a$ is the Wilson-Dirac operator with the negative mass term $\rho/a$, $a$ is the lattice spacing in physical units, $M$ is the Wilson hopping term with $r=1$.
Fermionic fields comply with periodic boundary conditions imposed on space and anti-periodic boundary conditions imposed on time.
The sign function is calculated by the  minmax polynomial approximation
\begin{equation}
D_W/ \sqrt{D^{\dagger} D_W}=\gamma_5 \sign(H_W),
\label{sign_function}
\end{equation}
where $H_W=\gamma_5 D_W$ is the hermitian Wilson-Dirac operator.
To compute the sign function we use the 50 lowest Wilson-Dirac eigenmodes.

The Dirac operator in  continuous space is $D=\gamma^{\mu} (\partial_{\mu}-iA_{\mu})$ and the corresponding Dirac equation looks as follows:
\begin{equation}
D \psi_k=i \lambda_k \psi_k.
\label{Dirac}
\end{equation}
The Neuberger overlap operator allows to calculate  eigenfunctions $\psi_k$
and  eigenvalues $\lambda_k$ for the test quark in external gauge field $A_{\mu}$.
$A_{\mu}$ is the sum of $SU(2)$ gauge field and the external abelian uniform magnetic field. The eigenmodes of the Dirac operator make it possible to construct operators and correlators presented in Section 3.

Abelian gauge fields interact only with quarks. To include  abelian magnetic field in the simulations we perform the
following substitution:
\begin{equation}
A_{\mu \, ij}\rightarrow A_{\mu \, ij} + A_{\mu}^{B} \delta_{ij},
\label{exchange}
\end{equation}

\begin{equation}
 A^B_{\mu}(x)=\frac{B}{2} (x_1 \delta_{\mu,2}-x_2\delta_{\mu,1}).
\end{equation}
To match this exchange with  lattice boundary conditions the twisted boundary conditions for fermions have been used \cite{Al-Hashimi:2009}.

Magnetic field in our calculations is directed along the third axes, its value  is quantized
\begin{equation}
qB=\frac{2\pi k}{(aL)^2}, \ \ k \in \mathbb{Z},
\label{quantization}
\end{equation}
where $q = - 1/3\, e$ is the electric charge of the $d$-quark. There is one type of fermions in the theory. The quantization condition imposes a limit on the minimal value of the magnetic field.
It equals to  $\sqrt{eB}=476.13\ \Mev$ for the lattice volume $18^4$ and  lattice spacing $a=0.0998\ \fm$. We are  far from the saturation regime in magnetic field when $k/(L^2)$ is not small because we use the values of $k$ in  the interval $0 - 6$ for $18^4$ lattice volume and $k=0 - 14$ for $14^4$ and $16^4$ lattice volumes.   For  inversion of the mentioned operator we use the Gaussian source (with the radius $r = 1.0$ in lattice units in both
spatial and time directions) and the point sink (a quark position  smeared over the Gaussian profile).

Our calculations were performed on symmetric lattices with the lattice volumes $14^4$, $16^6$, $18^4$  and  lattice spacings $a=0.0681, \ 0.0998$ and $0.1383\ \fm$.
We use statistically independent configurations of the gluon field  for  each value of the quark mass in the interval $m_q a = 0.01 - 0.8$. For quark masses larger than $0.01$ the inversion works well.
 \section{Observables}
 \label{Observables}
The following observables were calculated
\begin{equation}
\langle\psi^{\dagger}(x) O_1 \psi(x) \psi^{\dagger}(y) O_2 \psi(y)\rangle_A,
\label{observables}
\end{equation}
where $O_1, O_2=\gamma_5,\, \gamma_5 \gamma_{\mu},\, \gamma_{\mu}$ are Dirac gamma matrices, $\mu, \nu=1,..,4$ are Lorentz indices. In the Euclidean space $\psi^{\dagger}=\bar{\psi}$
~\cite{Vainshtein:1982}. In presence of gauge field being the sum of gluonic fields and abelian constant magnetic field these 
current-current correlators in a meson channel can be expressed via Dirac propagators.

In the continuous field theory \eq{observables} can be written as follows:
\begin{equation}
\int DA_{\mu} e^{-S_{YM}[A_{\mu}]} \left[ \Tr \left(\frac{1}{D+m} O_1 \right) \Tr \left( \frac{1}{D+m} O_2 \right)\right]-
\label{observables:cont}
\end{equation}
$$
\int DA_{\mu} e^{-S_{YM}}[A_{\mu}] \left[ \Tr \left( \frac{1}{D+m} O_1 \frac{1}{D+m} O_2 \right)   \right].
$$
The first term in the numerator of (\eq{observables:cont}) is the disconnected part, while the second one is  connected.
We checked that the disconnected part makes rather little relative  contribution to  correlators
 in comparison with  the connected part in our model. Also this disconnected part 
 does not effect the values of meson masses.
Therefore we calculate only the connected parts of  correlators~\eq{observables:cont}.

Correlators \eq{observables:cont} are defined by  Dirac propagators.
To calculate these correlators we should first determine the inverse matrix
for the massive Dirac operator.
For  the M lowest eigenstates of the Dirac operator this matrix is represented by the sum:
\begin{equation}
\frac{1}{D+m}(x,y)=\sum_{k<M}\frac{\psi_k(x) \psi^{\dagger}_k(y)}{i \lambda_k+m},
\label{lattice:propagator}
\end{equation}
where $M=50$.
Observables \eq{observables} have the following form
\begin{equation}
\langle \bar{\psi} O_1 \psi \bar{\psi} O_2 \psi \rangle_A=
\label{lattice:correlator}
\end{equation}
$$ \sum_{k,p<M} \frac{\langle k|O_1|k\rangle \langle p|O_2|p  \rangle-\langle p|O_1|k \rangle \langle k|O_2|p \rangle}{(i\lambda_k+m)(i\lambda_p+m)}
$$

The first term in \eq{observables:cont} is not essential as we mentioned above.
Magnetic and gluonic fields are considered in the way described above.

The mass of  neutral $\rho$ meson was extracted from the correlator of vector currents  $\langle j^V_{\mu}(x) j^V_{\nu}(y) \rangle_A$, where $j^V_{\mu}(x)=\psi^{\dagger}(x) \gamma_{\mu} \psi(x)$. We calculated the mass that has magnetic 
fieldwise projection of its spin equal to zero and  it corresponds to  $O_1,  O_2=\gamma_3$ in the expression
\eq{lattice:correlator}.
The mass of a neutral $A$ meson can be found  from the correlator of axial currents $\langle j^A_{\mu}(x) j^A_{\nu}(y) \rangle_A$, where $j^A_{\mu}(x)=\psi^{\dagger}(x) \gamma_5\gamma_{\mu}  \psi(x)$.
The correlator $\langle j^{PS}(x) j^{PS}(y) \rangle_A$ enables us to compute the mass of  $\pi$  meson, where
  $j^{PS}=\psi^{\dagger}(x) \gamma_{5} \psi(x) $ is the pseudoscalar current.
 \section{Methods}
 \label{Methods}
 To calculate masses we apply two methods.
The first one is based on spectral expansion of the lattice correlation function
$$
C(n_t)=\langle \psi^{\dagger}(\vec{0},n_t) O_1 \psi(\vec{0},n_t) \psi^{\dagger}(\vec{0},0) O_2 \psi(\vec{0},0)\rangle_A =
 $$
\begin{equation}
\sum_k\langle 0|O_1|k \rangle \langle k|O^{\dagger}_{2}|0 \rangle e^{-n_t a E_k},
\label{sum}
 \end{equation}
 \begin{equation}
C(n_t)= A_0 e^{-n_t a E_0} + A_1 e^{-n_t a E_1} + ... \ ,
 \label{sum2}
\end{equation}
where $A$ is some constant value, $E_0$ is the energy of the lowest state. For a particle with average momentum equal to zero $\langle\vec{p}\rangle=0$ this energy is equal to its mass $E_0=m_0$. $E_1$ is the energy of the first excited state, $a$ is the lattice spacing, $n_t$ is the number of a lattice site in the  line of  time direction. From  expansion \eq{sum2} one can see that for large values  $n_t$ the main contribution comes from the   ground state energy.

Due to periodic boundary conditions the contribution of the ground state into meson propagator  has the form
$$
C_{fit}(n_t)=A_0 e^{-n_t a  E_0} + A_0 e^{-(N_T-n_t)  a E_0}=
$$
\begin{equation}
2A_0 e^{-N_T a E_0/2} \cosh ((N_T-n_t) a E_0).
 \label{sum33}
\end{equation}
The  value of the ground state mass can be obtained by fitting the function \eq{sum33} to the lattice correlator \eq{sum}.
To minimize errors we take various  $n_t$ values from the  interval $4<n_t<N_T-4$.

 The second method  we use is the Maximal Entropy Method (MEM)~\cite{Asakawa:2001}.
 Euclidean correlator of the imaginary time $G(\tau, \vec{p})=\int d^3\,x \langle O(\tau, \vec{x}) O^{\dagger}(0,\vec{0})\rangle e^{-i\vec{p}\vec{x}}$ corresponds to the spectral function $\rho(\omega, \vec{p})$ as follows:
\begin{equation}
G(\tau,\vec{p})=\int_0^{\infty} \frac{d \omega}{2\pi} K(\tau, \omega) \rho(\omega, \vec{p}).
\label{integral}
\end{equation}

In general case  $ \rho(\omega, \vec{p})$  contains all the properties of mesons and hadrons having desired quantum numbers. We presume $\langle \vec{p}\rangle =0$ and do not consider any functions from it.
The first peak in the spectral function corresponds to  the energy of the ground state.
The kernel in~\eq{integral} can be expressed as follows:
\begin{equation}
K(\tau,\omega)=\frac{\cosh[\omega(\tau-1/2T)]}{\sinh(\omega/2T)},
\label{kernel}
\end{equation}
 where $T$ is the temperature, $\tau$ is the Euclidean time, $\omega$ is the  frequency.
 To extract the spectral function we should perform  an inversion of the equation \eq{integral}.

This problem on the lattice is ill-defined as
 the correlator $G(\tau)$ can be  calculated only  numerically at discrete  points $\tau_i=\tau_{min}+(i-1)a$, $i=1,...N_{\tau}$, and  $N_{\tau} \sim 10 - 50$.
The integral was approximated by the discrete sum at points $\omega_n=n \bigtriangleup \omega$,  $n=1,...,N_{\omega}$ and $N_{\omega}$ is usually $\sim O(10^3)$.
We cut  the integral~\eq{integral} off at some $\omega_{max}$.
 Nevertheless  this inversion turns out to be impossible.
Yet the ideas of Bayesian probability theory make it possible to overcome this problem.

The most probable spectral function $\rho(\omega)$ can be computed  provided that we find the maximum of  conditional probability $P[\rho|D H \alpha m]$, where $D$ is the data, $H$ is our hypothesis, $\alpha$ is a real and positive parameter, $m=m(\omega)$
is a default model.
This procedure is equivalent to the maximization of free energy
$F=L-\alpha S$, where $S$ is Shannon entropy,
\begin{equation}
S=\int_0^{\infty} d\omega \left[\rho(\omega)-m(\omega)-\rho(\omega)\ln\frac{\rho(\omega)}{m(\omega)}\right].
\label{entropy}
\end{equation}
$L$ is the standard likelihood function. Detailed explanation of how to make a corresponding discretization on the  lattice is offered  in \cite{Asakawa:2001}. 
We take $\alpha \in [\alpha_{min}, \alpha_{max}]$ and average the data within this interval.
This interval was chosen in such a way that the results may vary slightly (by  approximately $10 \%$).

The kernel~\eq{kernel} contains divergence  at $\omega=0$ leading to the unstable behaviour of the procedure under small energies.
The Bryan's key idea was to redefine the kernel  and the spectral function
\begin{equation}
\bar{K}(\omega,\tau)=\frac{\omega}{2T} K(\omega,\tau), \ \ \ \ \bar{\rho}(\omega)=\frac{2T}{\omega} \rho(\omega),
\label{newkernel}
\end{equation}
so that $K(\omega, \tau) \rho(\omega)=\bar{K}(\omega, \tau) \bar{\rho}(\omega)$ and apply the SVD theorem to the
modified discretized kernel $\bar{K}(\omega_n,\tau_i)$, see~\cite{Bryan:1990}.
We use this modified algorithm to determine the spectral function in the following form
\begin{equation}
\bar{\rho}(\omega)=\bar{m}(\omega) \exp \sum_{i=1}^N \bar{c}_i \bar{u}_i(\omega).
\label{spectralfunction}
\end{equation}

The column vectors $u_i,\ (i=1,..,N)$ are normalized
\begin{equation}
\langle u_i | u_j\rangle \equiv \sum_{n=1}^{N_{\omega}} u_i(\omega_n) u_j(\omega_n)=\delta_{ij},
\label{normalization}
\end{equation}
$c_i$ are the coefficients and we consider $\bar{K}(0, \tau)=1$.

Therefore, to reconstruct spectral function $\rho(\omega)$ we have to choose  default model
$\bar{m}(\omega)$ correctly.
Such a  default model should describe high and low energy behaviours of the spectral function correctly.
According to the analysis  \cite{Aarts:2007} we choose the following form of the model:
\begin{equation}
\bar{m}(\omega)=m_a\omega+m_b, \ \ m_a=\frac{G(N_{\tau}/2)}{ T^2}, \ \ m_b=a_H \frac{3}{8 \pi^2},
\label{defaultmodel}
\end{equation}
 where $a_H=1$ for scalar and pseudoscalar channels, $a_H=2$ for vector and axial vector channels \cite{Karsch:2001}.
 We also try to apply other default models (constant function, $\sim \omega^2$, vary the $m_a$ and $m_b$), but the choice \eq{defaultmodel} gives the best convergence for the MEM.

\section{Vector meson mass at $B=0$}

\label{ZeroField}
In this section we present the calculation of  neutral $\rho$ meson mass in $SU(2)$ gluodynamics under zero magnetic field. We compare our results to the previous ones to make sure that our method works properly.
We measure the correlator of  pseudoscalar currents $C^{PS PS}(n_t)=\langle j^{PS}(\vec{0}, n_t) j^{PS}(\vec{0},0) \rangle_A$, where
$j^{PS}(\vec{0}, n_t)=\bar{\psi}(\vec{0},n_t)\gamma_5 \psi(\vec{0},n_t)$ and calculate the mass of the lowest energy state of  neutral $\pi$ meson for different quark masses, volumes and lattice spacings.
\begin{figure}[htb]
\begin{center}
 \includegraphics[height=3in, angle=-90]{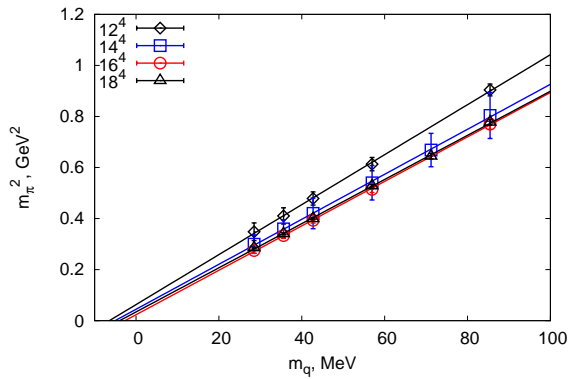}
\caption{The squared mass of the neutral $\pi$ meson calculated via the pseudoscalar correlator $C^{PSPS}(n_t)$ versus the bare lattice quark mass for  $12^4,14^4,16^4,18^4$ lattice volumes, $0.1383\ \fm$ lattice spacing, $\beta=3.1000$ and zero external magnetic field.}
 \label{fig:pion1}
\end{center}
\end{figure}
\begin{figure}[htb]
\begin{center}
 \includegraphics[height=3in, angle=-90]{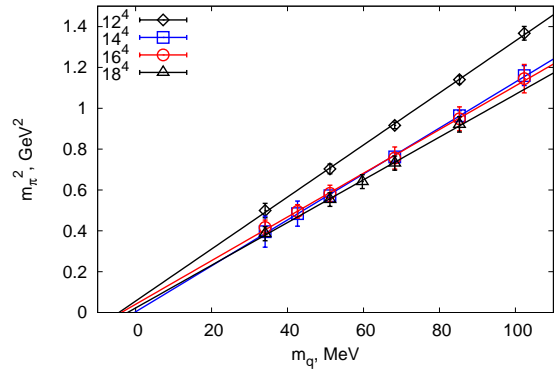}
\caption{The same function as on the Fig.\ref{fig:pion1}, but for the lattice spacing $0.1155\ \fm$ and $\beta=3.2000$.}
 \label{fig:pion2}
\end{center}
\end{figure}
Fig.\ref{fig:pion1} and \ref{fig:pion2} demonstrate the linear dependence of the squared $\pi$ meson mass versus the bare quark mass. The chiral perturbation theory predicts  linear dependence of squared pion mass on the renormalized quark mass $m^{ren}_q$, and this dependence can be expressed as follows:
\begin{equation}
f_{\pi}^2 m^2_{\pi}= m^{ren}_q \langle \bar{\psi} \psi \rangle,
\label{linear}
\end{equation}
where $f_{\pi}$ is the pion decay constant, $\langle \bar{\psi} \psi \rangle$ is the chiral condensate. 
Provided that the mass of quark tends to zero, the pions are massless.
 Fig.\ref{fig:pion1} and \ref{fig:pion2} show  that mentioned functions are slightly shifted relative to the origin of the coordinates and that the shift of quark mass from zero corresponds to the renormalization of this quark mass on the lattice.

\begin{figure}[htb]
\begin{center}
 \includegraphics[height=3in, angle=-90]{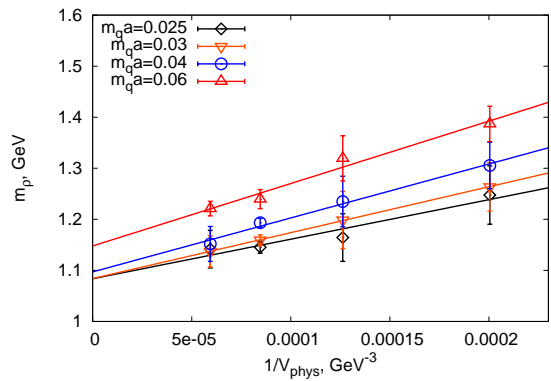}
\caption{The  extrapolation  of the neutral $\rho$ meson mass to the infinite physical volume under zero external magnetic field. Masses were calculated via the vector correlator $C^{VV}(n_t)$ for several bare quark masses, lattice volumes, lattice spacing $0.1383\ \fm$, $\beta=3.1000$, and $V_{phys}=(aL)^3$ is the physical volume.}
 \label{fig:rho1}
\end{center}
\end{figure}

\begin{figure}[htb]
\begin{center}
 \includegraphics[height=3in, angle=-90]{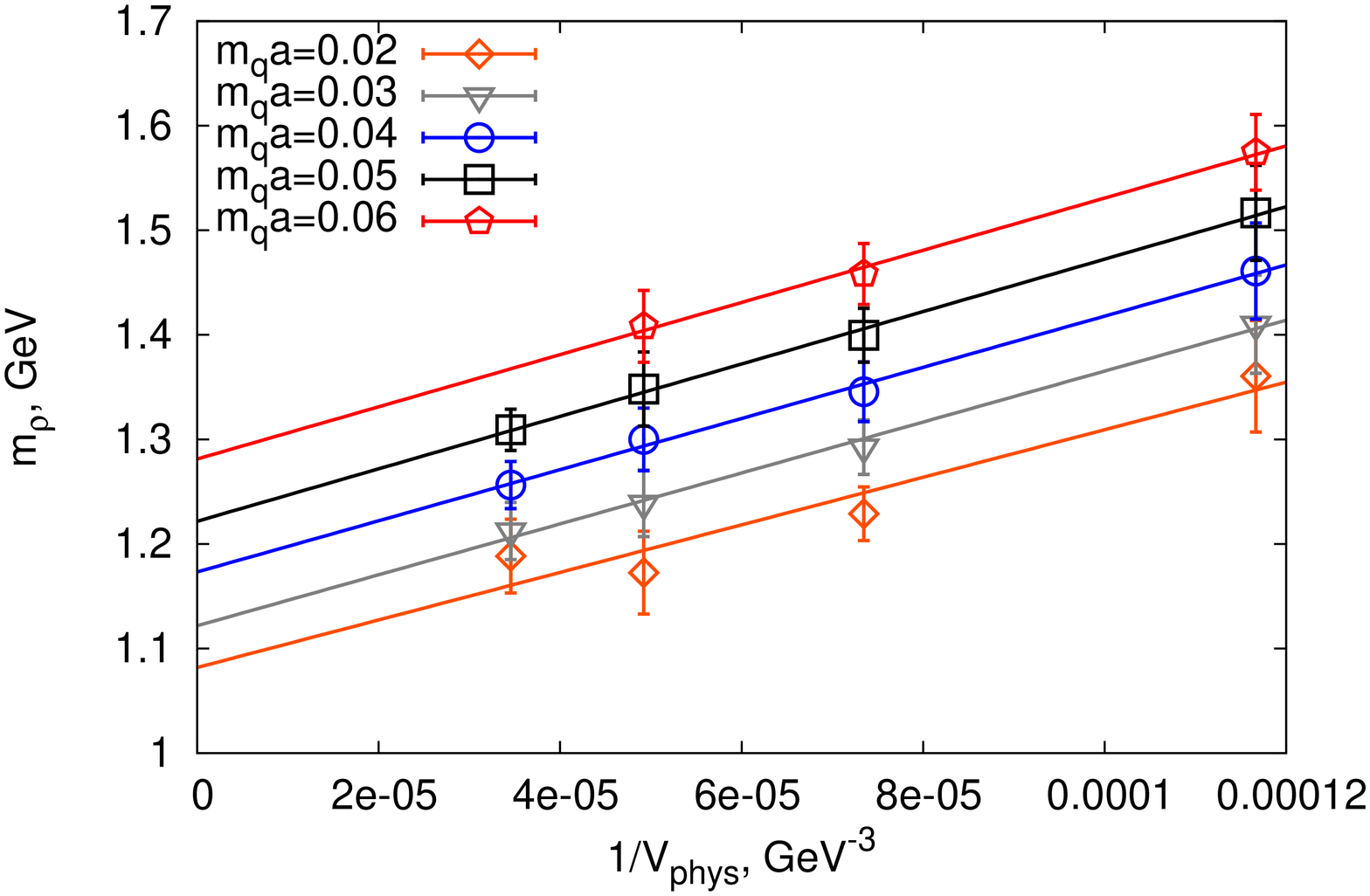}
\caption{The same function as on Fig.\ref{fig:rho1}, but for the  lattice spacing $0.1155\ \fm$ and $\beta=3.2000$. }
 \label{fig:rho2}
\end{center}
\end{figure}

We calculate the mass of  neutral vector $\rho$ meson under zero magnetic field via the correlators of vector currents. The symmetry between the different spatial directions was taken into account which improved the statistics significantly
(thrice).
The quark mass renormalization was not considered because its value is very small at zero field.
 The  extrapolation of $\rho$ meson mass to the limit of infinite physical volume  was performed for several values of  quark mass and two lattice spacings $0.1383\ \fm$ and $0.1155\ \fm$ (Fig.\ref{fig:rho1} and \ref{fig:rho2}).
\begin{figure}[htb]
\begin{center}
 \includegraphics[height=3in, angle=-90]{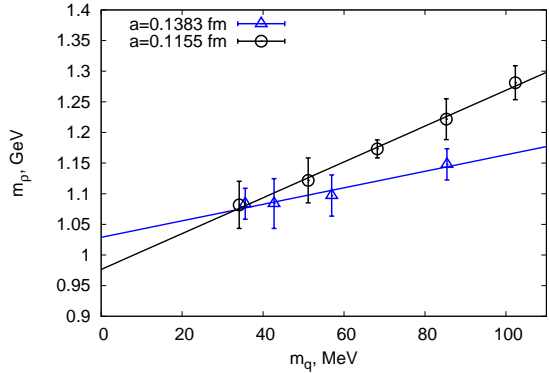}
\caption{The mass of the neutral vector $\rho$ meson for various quark masses and two lattice spacings. The  extrapolation was performed to the physical  mass of pion $m_{\pi}=135\ \Mev$. The results were obtained under zero external magnetic field by means of fitting.}
 \label{fig:rho3}
\end{center}
\end{figure}
After all we extrapolated $m_{\rho}$ to the quark mass $m_{q_0}$ corresponding to the  value of the $\pi$ meson mass equal to $135\ \Mev$. We obtained $m_{\rho}\simeq 980\pm 30\ \Mev$ for the lattice spacing $a=0.1338\ \fm$ and $1020 \pm 20\ \Mev$ for $a=0.1155\ \fm$ in SU(2) gluodynamics.
\section{The masses of mesons at $B\neq0$}
\label{Results1}
\begin{figure}[htb]
\begin{center}
 \includegraphics[height=3in, angle=-90]{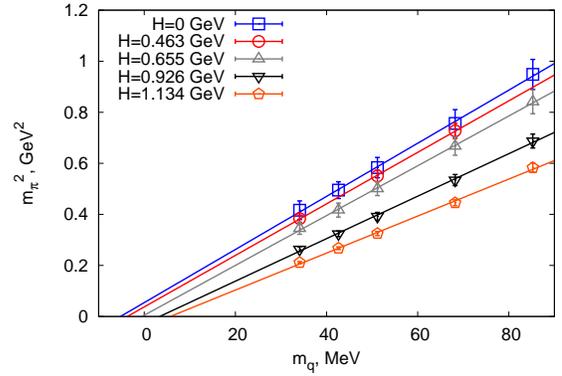}
\caption{The squared mass of the neutral $\pi$ meson calculated via  pseudoscalar correlator $C^{PSPS}(n_t)$ versus  bare lattice quark mass for the lattice volume $16^4$, lattice spacing $0.1155\ \fm$, $\beta=3.2000$ and diverse values of the external magnetic field.}
 \label{fig:pion}
\end{center}
\end{figure}
\begin{figure}[htb]
\begin{center}
 \includegraphics[height=3in, angle=-90]{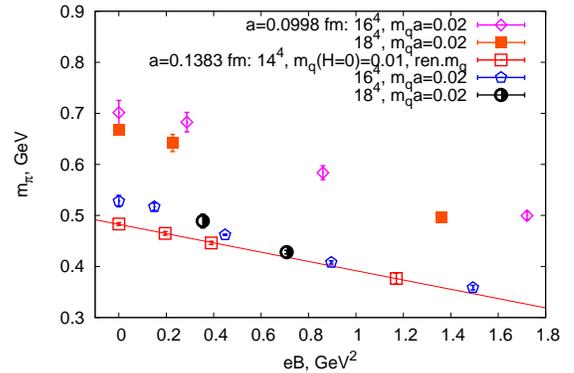}
\caption{The mass of the neutral $\pi$ meson obtained via the $C^{PSPS}(n_t)$ versus the squared value of the magnetic field for renormalized and nonrenormalized quark masses.}
 \label{fig:piondep}
\end{center}
\end{figure}
\begin{figure}[htb]
\begin{center}
\begin{tabular}{cc}
 \includegraphics[height=3in, angle=-90]{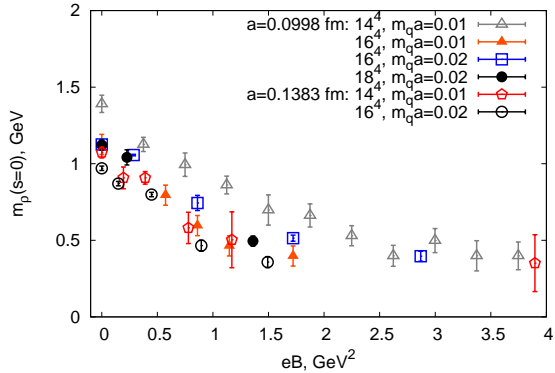}
 \end{tabular}
\caption{Dependence of the mass of the neutral vector $\rho$ meson with zero spin  $s=0$
 on the value of external magnetic field for the lattice volumes
 $14^4$, $16^4$, $18^4$ and lattice spacings $a=0.0998\ \fm, 0.1338\ \fm$ calculated in
 accordance with the Maximal Entropy Method.}
\label{fig:rho33_mem}
\end{center}
\end{figure}
\begin{figure}[htb]
\begin{center}
\begin{tabular}{cc}
 \includegraphics[height=3in, angle=-90]{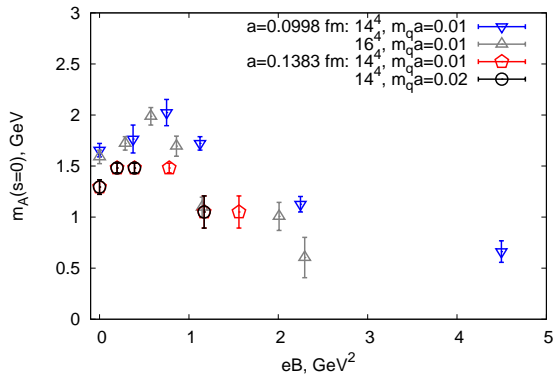}
\end{tabular}
\caption{The mass of the neutral axial $A$ meson with zero spin $s=0$
versus the value of external magnetic field for the lattice volumes
 $14^4$, $16^4$ and   lattice spacings $a=0.0998\ \fm, 0.1338\ \fm$ accordance with the Maximal Entropy Method.}
\label{fig:axialmem}
\end{center}
\end{figure}

On Fig.\ref{fig:pion} the squared pion mass is depicted for $16^4$ lattice volume, various bare quark masses and different values of the magnetic field $H=\sqrt{eB}$.
The observed linear dependence is legitimate for  the chiral perturbation theory.

Fig.\ref{fig:piondep} represents  the $\pi$  meson mass versus the value of the squared magnetic field.
The mass  depends on the lattice volume not significantly. For the renormalized quark mass and magnetic fields less than $1\ \Gev$ we have obtained linear  dependence of the mass on the magnetic field value; the slope of the graph of this function is negative, which agrees with results  of A.Smilga obtained in the chiral perturbation theory \cite{Smilga:1997}.
 The angle of the slope is obtuse and it differs from ChPT because we explore the  $SU(2)$ gauge theory without dynamical quarks. Thus $SU(2)$ reveals  qualitative properties of the theory correctly.

If the direction of external magnetic field is parallel to the 3rd coordinate axis then meson masses having magnetic fieldwise  projections of their spins equal to zero are calculated from the expression
\eq{lattice:correlator} with  $O_1,  O_2=\gamma_3$. The diagonal components of the correlators are not equal to zero
while the nondiagonal  ones equal to zero within the error bars.
Fig.\ref{fig:rho33_mem}  shows the mass of neutral vector meson with zero spin calculated in accordance with the Maximal Entropy Method for different lattice volumes, spacings and bare quark masses. 
We have not performed mass extrapolations and  renormalizations though.
We found that the mass decreases under raising magnetic field for  all the  sets of data. We use the ensemble of $O(10$) results of MEM procedure  and calculate  average and standard deviations  for the best set of MEM parameters. 
To calculate errors we took the $\omega$-discretization into account as well.

Fig.\ref{fig:axialmem} shows behaviour of the neutral $A$ meson mass with zero spin  depending on the external magnetic field. The mass decreases as well but under  $eB \sim 0.7\ \Gev^2$ we observe the peak which might be a lattice artifact.

 The lattice artefacts are not significant yet  we are restricted by the small lattice spacing which is not fine
 enough and cannot explore the masses under strong magnetic fields.
As we know the radius of the lowest Landau level
\begin{equation}
r_L=\frac{1}{\sqrt{eB}}.
\end{equation}
We may assume that lattice artefacts might appear under $r_L=a$. Therefore we make simple  estimations
and obtain the  maximal value of  magnetic field $eB=3.9\ \Gev^2$
for the spacing $a=0.0998\ \fm$,
$2.9\ \Gev^2$ for the $a=0.1155\ \fm$  and $2.0\ \Gev^2$ for  $0.1338\ \fm$ lattice spacing.
\section{Lattice extrapolations at $B\neq 0$}
\label{Results2}
\begin{figure}[htb]
\begin{center}
\begin{tabular}{cc}
 \includegraphics[height=3in, angle=-90]{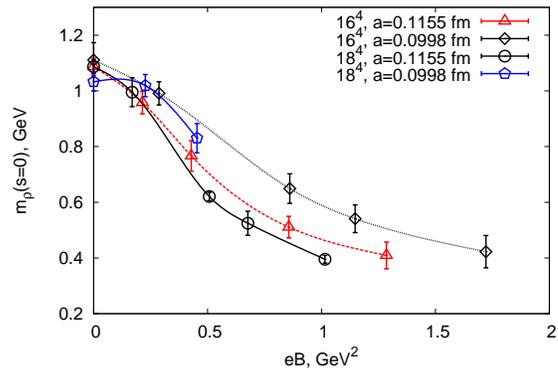}
\end{tabular}
\caption{Dependence of the mass of the neutral vector $\rho$ meson with spin   $s=0$  on the value of  external magnetic field for the lattice volumes $16^4$, $18^4$ and lattice spacings $a=0.0998,\ 0.1155\ \fm$.}
\label{fig:mrho_B2_extr_s0}
\end{center}
\end{figure}
\begin{figure}[htb]
\begin{center}
\begin{tabular}{cc}
 \includegraphics[height=3in, angle=-90]{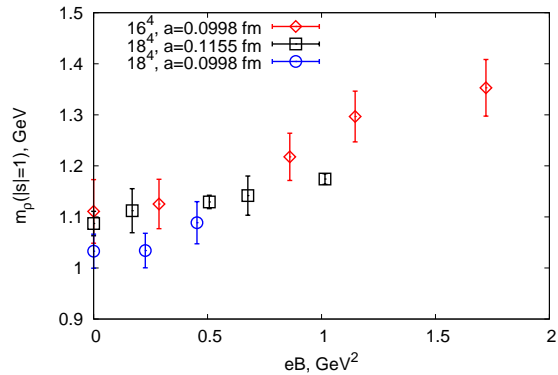}
\end{tabular}
\caption{The mass of the neutral vector $\rho$ meson with spin  $s=\pm 1$ versus the value of  external magnetic field for the lattice volumes $16^4$, $18^4$ and  lattice spacings $a=0.0998,\ 0.1155\ \fm$.}
\label{fig:mrho_B2_extr_s1}
\end{center}
\end{figure}
\begin{figure}[htb]
\begin{center}
\begin{tabular}{cc}
 \includegraphics[height=3in, angle=-90]{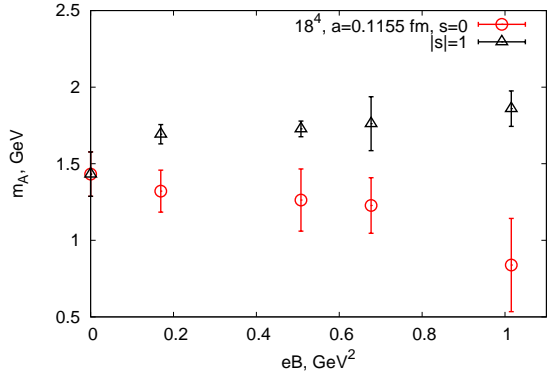}
\end{tabular}
\caption{The mass of the neutral axial $A$ meson with various spins versus the value of the magnetic field for the lattice volume $18^4$ and  lattice spacing $a=0.1155\ \fm$.}
\label{fig:mA_B2_extr_s0}
\end{center}
\end{figure}

On Fig.\ref{fig:mrho_B2_extr_s0}, \ref{fig:mrho_B2_extr_s1} and \ref{fig:mA_B2_extr_s0} we present the masses with various  spin projections 
that were received by fitting the coshinus function to correlators and further extrapolation of quark masses.

The masses of the  vector meson  were calculated for various values of  magnetic field.

We calculated these masses for nonzero magnetic field taking the quark mass renormalization $\delta m^{ren}_{latt}$ into account. On the basis of the mentioned calculation we can conclude that masses of vector mesons depend on that renormalization not significantly. Afterwards we performed $\rho$ meson mass extrapolation on the mass of quark to its value under which the mass of $\pi$ meson is equal to $135\ \Mev$. This procedure was done taking the renormalization of the quark mass into account.

 We calculate the mass of  $\rho$ meson  for several values of $m_q$ from the interval $m_q=0.01\div 0.8$, perform fitting and find coefficients $a_i$ and $b_i$ in the equations
\begin{equation}
m_{\rho}(s=0)=a_0+a_1 m_q,
 \label{fit2}
\end{equation}
\begin{equation}
m_A(s=0)=b_0+b_1 m_q.
 \label{fit3}
\end{equation}
Then we extrapolate $m_{\rho}(m_q)$ on physical values $m_{\rho}(m_{q_0})$  at $m_q=m_{q_0}$ using \eq{fit2} and \eq{fit3}.

Different components of the  vector currents correlators were calculated, and it was found that diagonal components 
differ from zero essentially while  nondiagonal ones are equal to zero within the error bars.
Correlators of the vector currents perpendicular  to the magnetic field are
$C_{11}^{VV}(n_t)=\langle j^V_1(\vec{0},n_t) j^V_1(\vec{0},0) \rangle_A$ and $C_{22}^{VV}(n_t)=\langle j^V_2(\vec{0},n_t) j^V_2(\vec{0},0) \rangle_A$, where $j^V_1(\vec{0},n_t)=\bar{\psi}(\vec{0},0) \gamma_1 \psi(\vec{0},n_t)$ and $j^V_2(\vec{0},n_t)=\bar{\psi}(\vec{0},0) \gamma_2 \psi(\vec{0},n_t)$.
 The masses with spin $s=\pm 1$ are found from
the relations $C^{VV}(s=1)=(C^{VV}_{11}+iC^{VV}_{22})/\sqrt{2}$ and $C^{VV}(s=-1)=-(C^{VV}_{11}-iC^{VV}_{22})/\sqrt{2}$.

On Fig.\ref{fig:mrho_B2_extr_s0} one can  see the dependence  of the mass of the  neutral vector $\rho$ meson with zero spin on the  value of the field after the mass renormalization and extrapolation for the lattice volumes  $16^4$ and $18^4$ and lattice spacings $a=0.0998,\ 0.1155\ \fm$. For the purposes of visualization we connected the points by splines.
The mass of vector meson decreases nonlinearly under raising magnetic  field for  all the lattices. We observe a weak dependence of masses from  lattice volumes and  spacings, but the qualitative behaviour for all the sets of data is the same.

  Fig.\ref{fig:mrho_B2_extr_s1} shows the dependence of the $\rho$ meson mass with nonzero spin on the field value. Masses with spin $s=\pm1$   increase under the raising  field.
These results were obtained after the quark mass extrapolation.

On Fig.\ref{fig:mA_B2_extr_s0} we see the mass of the neutral axial  meson with fieldwise   spin projections $s=0,\pm 1$.
Calculation of the axial meson mass requires  much more statistics than the calculation of the  vector meson mass
especially  in cases with nonzero spin components. One can observe that the mass of $A$ meson with zero spin decreases while the masses with $s=\pm 1$ increase slowly.

Unfortunately the quantum numbers of mesons on the lattice in the presence of  magnetic field  are not precise. The mixing takes place due to the interaction between photons and vector (axial) quark currents and can occur between neutral pion and neutral $\rho$ (or $A$) meson with zero spin. Now there are  no rigorous methods  to disentangle these two states in  magnetic field, this is the topic for the further work.
However, we have clear signs of the increase of the vector and axial mesons masses with $s=\pm 1$  in our $SU(2)$ theory.

\section{Conclusions}

In this work we explore the behaviour of masses of the neutral pseudoscalar  $\pi$, vector $\rho$ and axial $A$ mesons
in confinement phase in the presence of  external magnetic field of  hadronic scale.
 We observe that the masses with fieldwise spin projection equal to zero
  differ from the masses with spin projection $s=\pm 1$.
The masses with $s=0$ decrease with the growth of the magnetic field while the masses with $s=\pm 1$ increase
in the same conditions.
 We consider this phenomena to be the result of the anisotropy created by the  strong magnetic field.
We do not observe any condensation of neutral mesons, so there are no evidences of superfluidity in the confinement phase.
However, the presence of superconducting phase at high values of the magnetic field $B$~\cite{Chernodub:2010} in QCD is a hot topic for discussions. Condensation of charged $\rho$ mesons would be an evidence of the 	
existence of superconductivity in QCD.

\section{Acknowledgments}

The authors are grateful to  ITEP supercomputer center (the calculations were performed on supercomputers  "Graphin" and "Stakan") and Moscow Supercomputer JSCC Center. Also  we would like to express our deep appreciation to M.I.Polikarpov, M.N.Chernodub for their comments and advices.

\end{document}